\documentclass{aa}
\usepackage{graphicx}
\usepackage{subfigure}
\usepackage{txfonts}
\begin{document}
   \title{Synchrotron flaring behaviour of Cygnus~X-3 during the February--March 1994 and September 2001 outbursts}
   
         \author{E. J. Lindfors
          \inst{1,2}
	  \and
          M. T\"urler
	  \inst{3,4}
	  \and
	  D. C. Hannikainen
	  \inst{5}
	  \and
	  G. Pooley
	  \inst{6}
	  \and
	  J. Tammi
	  \inst{7}
	  \and
	  S. A. Trushkin
	  \inst{8}
          \and
	  E. Valtaoja
          \inst{1,9}
	  \fnmsep
          }

   \offprints{E.J. Lindfors email:elilin@utu.fi}

   \institute{Tuorla Observatory, V\"ais\"al\"a Institute of Space Physics and Astronomy, University of Turku, 21500 Piikki\"o, Finland
          \and
	      Mets\"ahovi Radio Observatory, Helsinki University of Technology, 02540 Kylm\"al\"a, Finland
          \and
             Geneva Observatory, University of Geneva, ch. des Maillettes 51, 1290 Sauverny, Switzerland
         \and 
             INTEGRAL Science Data Centre, ch. d'Ecogia 16, CH-1290 Versoix, Switzerland   
         \and
	     Observatory, PO Box 14, 00014 University of Helsinki, Finland
	 \and
	     Astrophysics, Cavendish Laboratory, J. J. Thomson Avenue, Cambridge CB3 0HE
         \and
             UCD School of Mathematical Sciences, University College Dublin,
Belfield, Dublin 4, Ireland
         \and 
             Special astrophysical observatory RAS, Nizhnij Arkhyz, 369167, Russia
	 \and
	     Department of Physics, University of Turku, 20100 Turku, Finland
}

   \date{Received, accepted}

   \abstract {} 
{In this paper we study whether the shock-in-jet
model, widely used to explain the outbursting behaviour of quasars,
can be used to explain the radio flaring behaviour of the microquasar
Cygnus~X-3.  }  
{We have used a method developed to model the
synchrotron outbursts of quasar jets, which decomposes multifrequency
lightcurves into a series of outbursts. The method is based on the
Marscher \& Gear (1985) shock model, but we have implemented the
modifications to the model suggested by Bj\"ornsson \& Aslaksen
(2000), which make the flux density increase in the initial phase less
abrupt. We study the average outburst evolution as well as specific
characteristics of individual outbursts and physical jet properties of
Cyg~X-3.}  
{We find that the lightcurves of the February--March 1994
and September 2001 outbursts can be described with the modified
shock model. The average evolution shows that instead of the expected
synchrotron plateau, the flux density is still increasing during the
synchrotron stage.  We also find that high frequency peaking outbursts
are shorter in duration than the ones peaking at lower frequencies.
Finally, we show that the method can be used, complementary to
radio interferometric jet imaging, for deriving the physical parameters such as the
magnetic field strength and the energy density of relativistic electrons in
the jet of Cyg~X-3.}  
{} 

   \titlerunning{Synchrotron flaring behaviour of Cyg~X-3}
   \maketitle
%

\section{Introduction}
X-ray binary systems exhibiting jets with relativistic motions are
called microquasars. They are generally regarded as the galactic
counterparts of more powerful extragalactic sources, quasars. This
suggests that the underlying physical processes that operate in these
systems are the same and thus are manifested in their radiative
processes.

Cygnus~X-3 is one of the most intensively studied microquasars. It is
a strong X-ray source, and is thought to consist of a compact object
accreting matter from a Wolf-Rayet star (van Kerkwijk et
al. 1996). The source occasionally undergoes huge radio outbursts in
which the flux density can increase up to levels of $\sim 20$ Jy at
frequencies of a few GHz. The radio flares more typically consist of a
few peaks of 1-5\,Jy and are preceded by a quenched state (Waltman et
al.  1994). During recent outbursts, jet-like structures have been
observed at radio frequencies.
On milliarcsecond scales a one-sided jet was observed with the Very Long
Baseline Array (VLBA) during one outburst of February 1997
(Mioduszewski et al. 2001, hereafter M01) while in September 2001 a two-sided jet was
observed (Miller-Jones et al. 2004, hereafter MJ04). On arcsecond scales a two-sided
jet has been observed with the Very Large Array (VLA) in September
2000 (Mart\'i et al. 2001).

The synchrotron emission of microquasars has historically been interpreted as 
originating from distinct expanding clouds of relativistic plasma (van
der Laan 1966, Hjellming \& Johnston 1988). More recently, the internal shock model -- originally developed for $\gamma$-ray bursts -- has been suggested to explain the observed variability of microquasars (Kaiser et al. 2000).
On the other hand, the model of Marscher \& Gear
(1985) -- describing analytically the synchrotron emission of a shock
wave propagating downstream in a relativistic jet -- has been the
baseline model for the flaring behaviour of quasars in the
radio-to-submillimetre range. Recently, this shock wave interpretation
has been shown to be able to describe well the observed lightcurves of
two microquasars: GRO~J1655-40 (Hannikainen et al. 2000, Stevens et
al. 2003) and GRS~1915+105 (T\"urler et al. 2004).

In this work we study a third galactic source, Cyg~X-3, showing
repeated outbursts with a typical evolution from high- to
low-frequencies as expected by the model of Marscher \& Gear (1985) but
with significantly longer outbursts than in GRS~1915+105. We use a
generalization of the shock-in-the-jet model of Marscher \& Gear (1985)
to describe two distinct flaring epochs of Cyg X-3: February--March 1994
and September 2001. We decompose simultaneously the multifrequency
lightcurves into a series of outbursts using a similar
methodology as for 3C~273 (T\"urler et al. 1999, 2000),
GRS~1915+105 (T\"urler et al. 2004) and 3C~279 (Lindfors et al. 2005,
2006). Preliminary results on the February--March 1994 period were
already published in Lindfors \& T\"urler (2007).

This method allows us to derive the observational and physical properties
of an average outburst in our data set. We then use this result to further
probe the physical conditions of the emission region in the jet of Cyg X-3.

\section{Data}
\begin{figure*}
\sidecaption \centering \includegraphics[width=12cm]{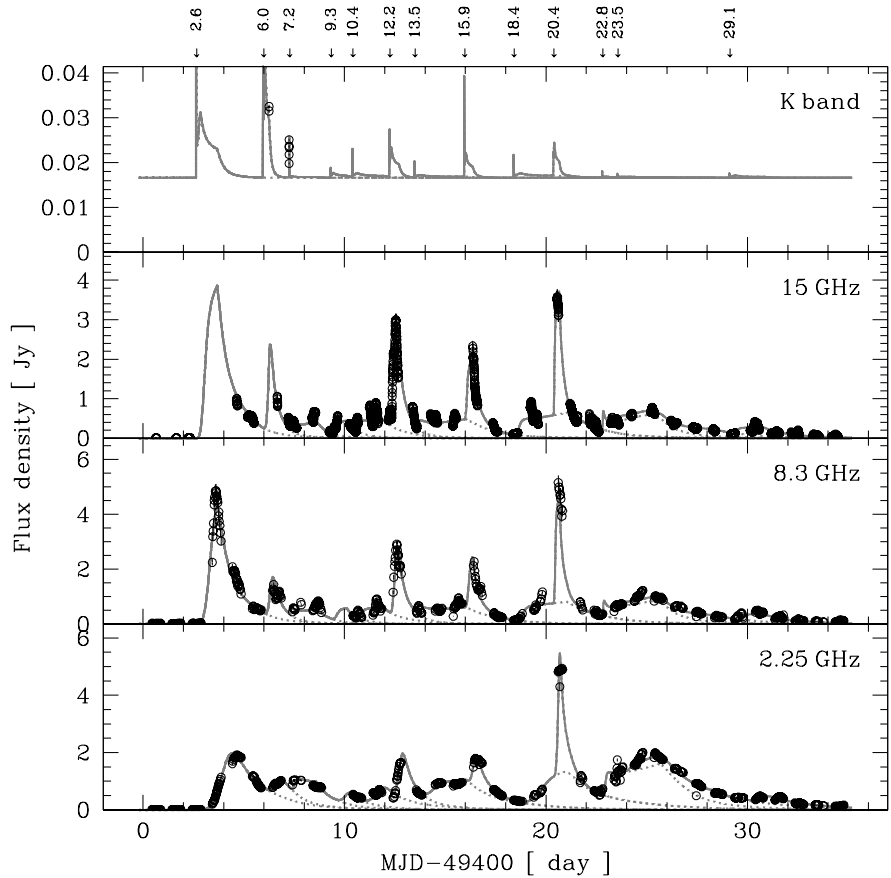}
\caption{Lightcurves of the February--March 1994 outburst decomposed
  into a series of 13 outbursts. The solid line shows the sum of all
  components and dashed lines separate the contribution of individual outbursts.
  The radio lightcurves are from the Ryle telescope (15~GHz) and the GBI
(2.25~GHz and 8.3~GHz) while the infrared data are from the UKIRT.
  Any additional constant quiescent flux is assumed to be negligible
except for the K-band lightcurve, where we fix it to 16.6\,mJy according
to Fender et al. (1996).
The time labels at the top of the figure show the starting dates of the outbursts.}
\label{FigVisu94}%
\end{figure*}

\begin{figure*}
\centering \includegraphics[width=\hsize]{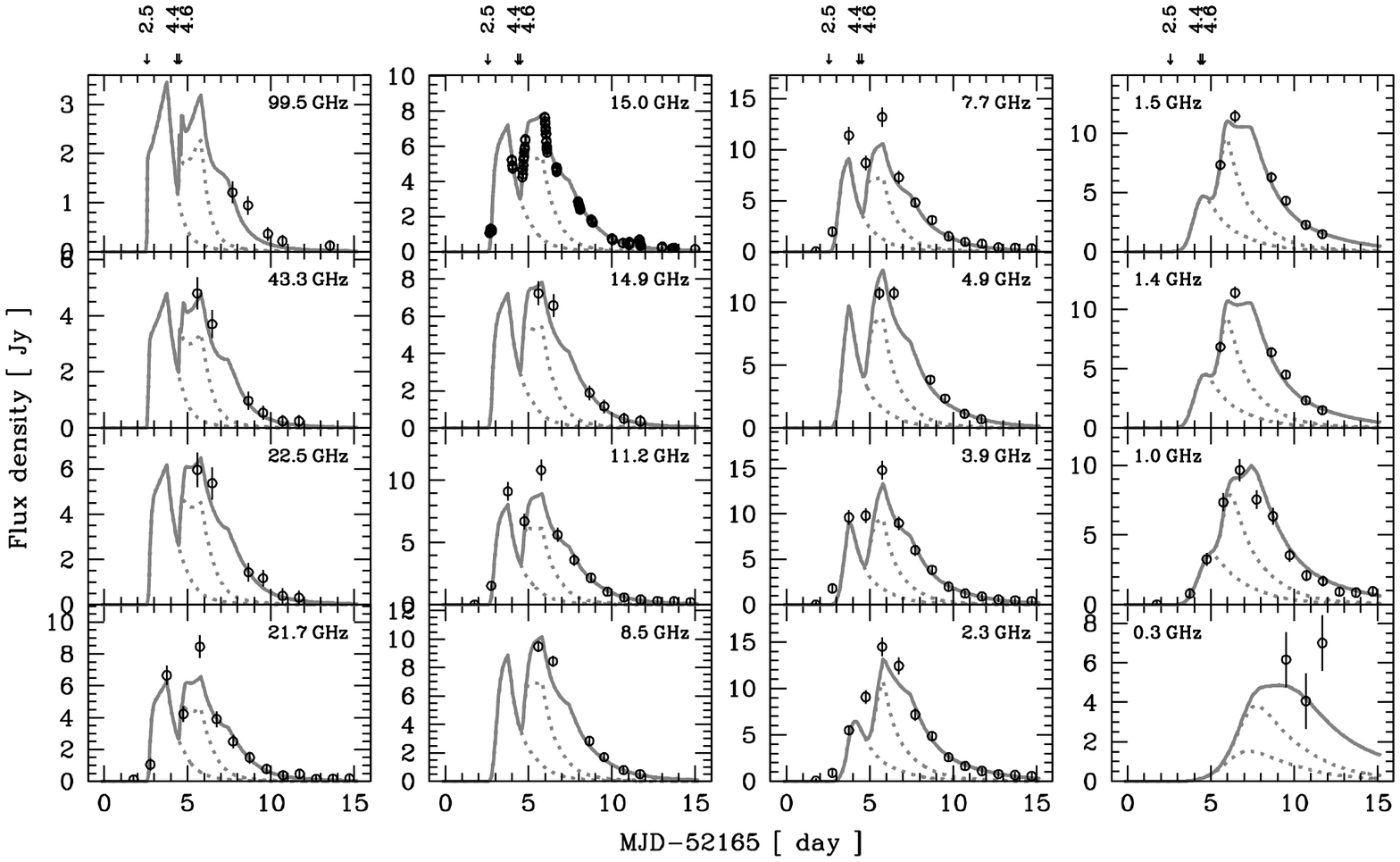}
\caption{Sixteen lightcurves of the September 2001 flaring period
decomposed into a series of three distinct outbursts. The solid line
shows the sum of the model outbursts and dashed lines separate
the contribution from individual events. The time labels at
the top of the figure show the starting dates of the outbursts. The data are from VLA (0.3275, 1.385, 1.465, 4.86, 8.46, 14.94, 22.46,
43.34\,GHz), RATAN 600 (1.0, 2.3, 3.9, 7.7, 11.2 and 21.7\,GHz), Ryle Telescope
(15\,GHz) and OVRO (99.48\,GHz).}
\label{FigVisu01}%
\end{figure*}

In February--March 1994 Cygnus~X-3 was observed to be flaring at radio
frequencies. The flaring lasted for approximately 30 days during which
time the radio flux density increased and decreased several times. The
maximum flux density of this flaring epoch was $\sim 5$ Jy at
2.25~GHz.  The dataset used here has been published by Fender et
al. (1997). It contains radio lightcurves from the Ryle telescope
at 15~GHz and from the Green Bank Interferometer (GBI) at 2.25~GHz and 8.3~GHz and
infrared data from the United Kingdom Infrared Telescope (UKIRT). A
possible constant emission in Cyg X-3 during quiescence is much
smaller than the peaks of the flares and hence can be considered
negligible in the radio, but not in the infrared. We therefore assume
a quiescent flux density contribution in the K-band of 16.6\,mJy as
measured on 7 August 1984 (Fender et al. 1996).

In September 2001 Cyg~X-3 underwent a major flare with flux densities
reaching 15\,Jy at 4\,GHz. This flaring period was quite
different from the one of February--March 1994: the maximum
flux density is much higher and judging from the lightcurves it seems
that it consists of only 2 to 3 distinct outbursts. For this flaring
period we use data from 16 frequencies. They are composed of the Very Large
Array (VLA, 0.3275, 1.385, 1.465, 4.86, 8.46, 14.94, 22.46, 43.34\,GHz)
and Owens Valley Radio Observatory (OVRO, 99.48\,GHz) data published by
MJ04, the RATAN 600 monitoring data at 1.0, 2.3,
3.9, 7.7, 11.2 and 21.7\,GHz 
, and the 15\,GHz lightcurve from the Ryle telescope. 
We noticed some calibration discrepancy between the
VLA data and the RATAN 600 and Ryle telescope data. We therefore
multiplied the VLA flux densities and errors by a factor 0.8 to match
the flux density scale of the other datasets. 

As our method consists in a $\chi^2$ fit of all data points simultaneously
(see Sect.~3), data with smaller uncertainties will have more weight in the
fit. If some lightcurves have much larger error-bars than others, they might
not have enough weight to well constrain the model. It
is therefore important to have lightcurves with comparable relative errors
to get an equilibrated fit across the frequency range. We have, therefore,
doubled the error estimates at 2.25\,GHz for the 1994 dataset. The original
uncertainties in the 2001 dataset are directly proportional to the flux. This
results in very small errors for low-flux measurements and thus gives to those
observations much more weight in the fit. To partly compensate for this effect
we add systematic errors of 0.3\,Jy to all data in the 1--22.5\,GHz frequency
range, 0.2\,Jy at 43.3\,GHz, and 0.1\,Jy at 99.5\,GHz.

\section{Model and Method}

The model we use to study the flaring behaviour of Cyg~X-3 is based on
the shock model of Marscher \& Gear (1985). This model considers
a disturbance - pressure or bulk velocity increase -- at the base of
the jet that will eventually become supersonic because of the pressure
gradient along the jet.  The resulting shock wave will accelerate
particules crossing the shock front that will emit an outburst of
synchrotron radiation in the increased magnetic field of the shocked
plasma. Marscher \& Gear (1985) identify three distinct phases during
the downstream propagation of a shock wave in a jet and describe
analytically the evolution of the associated synchrotron emission. In
the initial growth phase, when the Compton losses predominate, the
synchrotron self-absorption turnover frequency decreases and the
turnover flux density increases. In the second phase the synchrotron
losses dominate and the turnover frequency decreases while the
turnover flux density remains roughly constant.  The third phase is
the decay phase, when the adiabatic losses dominate and both the
turnover frequency and flux density decrease. Bj\"ornsson \& Aslaksen
(2000) proposed a modification to the initial Compton-loss stage
making the initial rise in flux density much less steep than in the
original calculation of Marscher \& Gear (1985). The model used here
takes this modification into account.

\begin{table*}
\begin{center}
\begin{tabular}{l l l l}\hline \hline
par.& description& value 1994& value 2001\\ \hline
$t_{\mathrm{r}}$& start time of the synchrotron stage& 0.12~d& 0.19~d\\
$t_{\mathrm{p}}$& time of peak flux density (start of adiabatic stage)& 1.15~d& 1.74d\\
$\nu_{\mathrm{p}}$& frequency of peak flux density& 1.75~GHz& 1.86~GHz\\
$S_{\mathrm{p}}$& peak flux density& 1.93~Jy& 10.74~Jy\\
$\nu_{\mathrm{b}}$& frequency of spectral break at time $t_{\mathrm{p}}$& 115~GHz& 122~GHz\\
$t_{\mathrm{f}}$& start time of optically thin slope flattening& 0.03~d& 0.04~d\\
$s$& index of electron energy distribution $N(E)\propto KE^{-s}$& 1.77\\
$k$& index of electron normalization  change $K\propto R^{-k}$  & 2.54\\
$r$& index of jet opening radius change with distance $R\propto L^r$& 1.21\\
$b$& index of magnetic field strength change $B\propto R^{-b}$& 1.57\\
$d$& index of Doppler factor change $D\propto R^{-d}$& 0.0\\ \hline
\end{tabular}
\caption{Physical and observational parameters describing an average outburst
during the 1994 and 2001 flaring periods. The parameter $d$ was fixed to
zero in our model. The ten remaining parameters have only been adjusted to the
1994 dataset, since we assume the same typical outburst evolution for the
2001 dataset (see text). We however give here the observational parameter values that
we derive on average in 2001 according to the specificities of the 3 outbursts given
in Table 2.}
\end{center}
\end{table*}

The generalized equations of the Marscher \& Gear (1985) shock
model were derived by T\"urler et al. (2000). This generalization
allows one to describe the emission from shocks which are not
necessarily propagating with constant velocity in a straight, conical
and adiabatic jet. The synchrotron outburst is characterized by the
frequency $\nu_{\mathrm{m}}$ and flux density $S_{\mathrm{m}}$ of the
synchrotron self-absorption turnover evolving with time. As the jet
opening radius $R\propto L^{r}$ is widening with distance along the
jet $L$ and thus with time $t$ since the onset of the shock wave, the
evolution of the observable quantities $\nu_{\mathrm{m}}$ and
$S_{\mathrm{m}}$ is defined by the dependency on $R$ of physical jet
quantities. These physical quantities are the normalization $K\propto
R^{-k}$ of the electron energy distribution $N(E)\propto KE^{-s}$, the
magnetic field strength $B\propto R^{-b}$ and the Doppler factor of
the emission region $D\propto R^{-d}$. In the model used here for
Cyg~X-3 we do not allow for a change of the Doppler factor by fixing
$d=0$.

To apply this model to the observations we use the methodology originally
proposed by T\"urler et al. (1999, 2000). It consists in fitting iteratively
subsets of all model parameters simultaneously to the complete observational
dataset. The model parameters describe fully the synchrotron
emission of a typical model outburst with a set of both physical and
observational parameters (listed in Table 1) from which individual
outbursts are only allowed to differ by their start time and by three additional
parameters. These three outburst specific parameters can either define a scaling
in time $t$, frequency $\nu$ and flux $S$ (T\"urler et al. 1999) or be the
consequence of a more physical scaling of the quantities $K$, $B$ and $D$
defined above (T\"urler et al. 2000). We choose here for Cyg~X-3 the more
phenomenological approach also used recently for 3C~279 by Lindfors et al.
(2006).



The model parameterization is simplified in Cyg~X-3, because of the
negligible underlying radio jet emission and the absence of a decaying
outburst at the start of the flaring periods.
We tested a model
accounting for the emission from the receding jet, as was used for
GRS~1915+105 (T\"urler et al. 2004), but this did not improve the fit
and therefore we decided to use a simple model where all the emission
originates from the approaching jet.
The jet velocity $\beta=v/c$ and orientation $\theta$ to the line of sight
derived from radio imaging observations (M01, MJ04) suggests that the condition
$\beta\approx \cos(\theta)$ for having a shock viewed sideways is not satisfied
(Marscher et al. 1992). We therefore use here the usual equations describing the
evolution of a shock viewed face-on to the observer. Other differences compared
to previous studies are that we do not impose here a linear decrease of the
magnetic field $B$ with the jet opening radius $R$ as $B\propto R^{-1}$, and we
do not include a spectral break below the synchrotron self-absorption
turnover frequency. Indeed, we found that the lack of data in the
submillimetre range, does not allow to well constrain this additional parameter
during the early phase of the shock evolution. The optically thick spectral
index is therefore fixed to 2.5 which is the theoretical value for a homogeneous
synchrotron source.

\section{Results and Discussion}
We have decomposed the lightcurves of the February--March 1994 and
September 2001 periods into a series of outbursts. The model for the
February--March 1994 data has a total of 62 free parameters: 10
parameters are used to describe the shape and evolution of the
synchrotron outburst (see Table~1), while the remaining 52 parameters
are used to define the start times of the 13 outbursts as well as
their specificity (see Table~2). The model fit to the 1994 lightcurves
is shown in Fig.~1. The September 2001 dataset is rather sparsely
sampled, and although it contains 16 frequencies, it has more than 10
times less data points than the 1994 dataset and therefore can only
loosely constrain the model. We chose to fix the average evolution and
jet parameters for the 2001 outbursts to those derived for the 1994
observations. The fit to the 2001 dataset has, therefore, only 12 free
parameters, which are the start times and specificities of the three
outbursts given in Table~2. The best fit model is shown for all 16
lightcurves in Fig.~2.

We obtain reduced $\chi^2$-values of $\chi^2$/d.o.f. = $\chi^2$/3314 =
10.2 and $\chi^2$/d.o.f. = $\chi^2$/261 = 1.80, for the 1994 and 2001
datasets, respectively.  These values, above unity, reflect the fact
that our fit is not able to fully describe any single data point. We
however consider these values as adequate, because the prime goal of
this study is not to have a perfect fit, but to derive the main
physical properties of the jet based on the available observations.
The model is therefore kept as simple as possible, in particular by
assuming self-similar outbursts in a jet with constant physical
properties. Furthermore, we do not consider here jet precession and
counter-jet emission.

Visual inspection of the fit reveals that the model curve remains
sometimes below the peaks of the observed lightcurves. This was also
found for 3C~279 and is a known problem of shock-in-jet models -- the
model seems not able to reproduce well the sharp peaks of the observed
lightcurves. However, this probelm could partly be due to the fact
that errors are often proportional to flux so that the $\chi^2$ fit
gives less weight to high-flux measurements compared to low-flux
observations.

\subsection{The Average Evolution of the Outbursts and Properties of Individual Outbursts}

The overall evolution of an average model outburst in Cyg~X-3 in
February--March 1994 is shown in Fig~3 and corresponding parameter
values are given in Table~1. The flux density increase during the
Compton stage is rather slow but during the synchrotron stage the
increase in flux density becomes more abrupt. The slow rise in the
flux during the Compton stage is a consequence of the use of the
modified outburst evolution for this phase proposed by Bj\"ornsson \&
Aslaksen (2000).  The existence of this Compton stage and the location
of the transition to the synchrotron stage are however only very
loosely determined by the few infrared measurements. A direct
transition to the adiabatic stage either from the synchrotron or the
Compton stage, which are difficult to distangle, is therefore not
excluded.

\begin{figure}
\centering 
\includegraphics[width=\hsize]{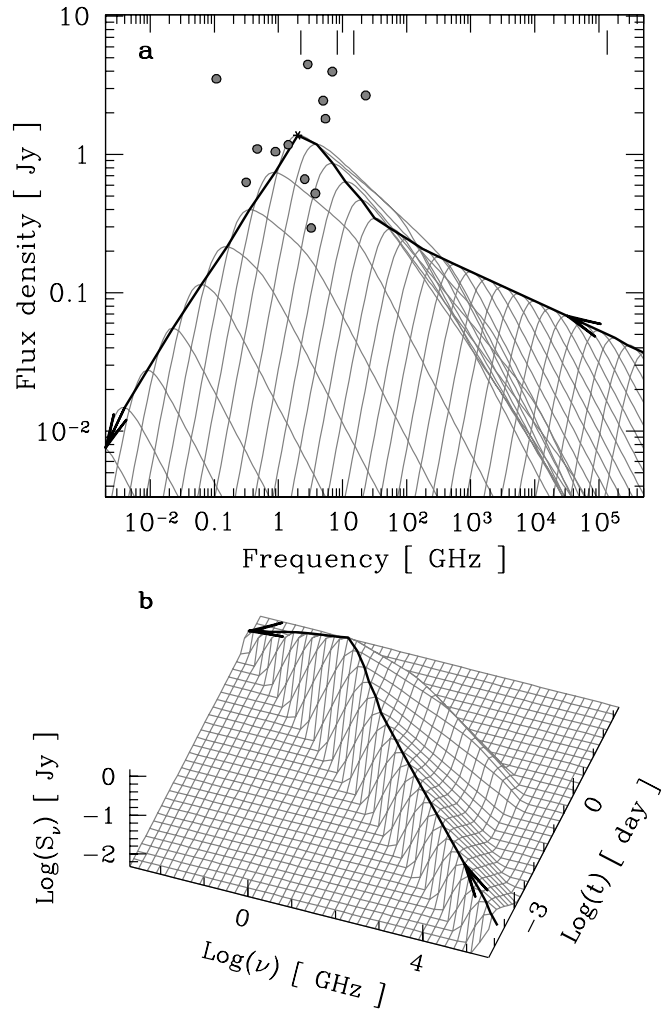}
\caption{a) Series of model spectra at different times during the evolution of
the average outburst. The filled circles refer to specific characteristics
of individual outbursts of the February--March 1994 as listed in Table~2. 
b) The three dimensional evolution of the average outburst in logarithmic
scale. The solid line with arrows shows the path followed by the spectral
self-absorption turnover.}
\label{Evo}%
\end{figure}

Individual outbursts differ from the average outburst by a rescaling
in flux ($S$), frequency ($\nu$) and time ($t$). This rescaling is
parameterized in the model by the logarithmic shifts given in
Table~2. These shifts change the amplitude $S_p$, the peaking
frequency $\nu_p$ and the duration $t_p$ of individual outbursts as
shown by the filled circles in Fig.~3.  To investigate whether the
shifts are correlated, we applied the Spearman rank-order test
(e.g. Press et al. 1992). For the February--March 1994 outburst, we
find a significant anti-correlation (likelyhood for such correlation
to appear by chance is less than 0.25\%) between the peaking frequency
and the duration, the highest peaking outbursts being shorter. The
same behaviour was previously found in 3C~273 (T\"urler et
al. 1999). As suggested by these authors this anti-correlation might
be related to the distance from the core at which the shock forms, the
short-lived and high-frequency peaking flares forming closer to the
core than longer-living low-frequency peaking outbursts.

Because the 2001 flaring period consists
of three outburst components only, any correlation study is not
applicable there. Compared to the 1994 period, the outbursts of 2001
differ mainly in amplitude (flux density), while timescales and frequency
ranges are similar (see Tables~1 and 2).

\begin{table}
\begin{center}
\begin{tabular}{l r r r r}\hline \hline
dataset&$T_0$&$\Delta$ log$S$&$\Delta$ log$\nu$&$\Delta$ log$t$\\ \hline
1994& 2.6 & 0.46&  0.54& -0.03 \\ 
 & 6.0 & 0.29& 1.06& -0.57 \\
 & 7.2 & -0.10&-0.63&   0.07 \\
 & 9.3 &-0.42& 0.28& -0.15  \\
 & 10.4& -0.32&  0.11&   0.13 \\ 
 & 12.2&  0.25&  0.40&  -0.38  \\
 & 13.5& -0.12& -0.35&   0.36\\
 & 15.9&  0.12&  0.43&  -0.43\\
 & 18.4& -0.07& -0.15&   0.37\\
 & 20.4&  0.51&  0.16&  -0.59\\
 & 22.8&  0.41& -1.27&   0.41\\
 & 23.5& -0.34& -0.80&   0.62\\
 & 29.1& -0.67&  0.21&   0.18\\ \hline
2001& 2.5& 0.81&  0.32&  0.05\\
 & 4.4& 0.76& -0.03&  0.09\\
 & 4.6& 0.67& -0.21&  0.41\\ \hline
\end{tabular}
\caption{The start times ($T_0$) of the model outbursts and their
  logarithmic shifts from the average shock evolution in flux density
  ($\Delta$ log $S$), frequency ($\Delta$ log $\nu$, negative if the
  outburst peaks at lower frequencies than the average outburst) and
  time scale ($\Delta$ log $t$, negative if the outburst evolution is
  faster than the evolution of the average outburst). The times $T_0$
  are expressed in days starting from 17 Feb.  1994 (MJD-49400) and
  from 13 Sep. 2001 (MJD-52165) for the datasets of 1994 and 2001,
  respectively. The shifts of the September 2001 dataset are given
  with respect to the average model outburst derived for
  February--March 1994 dataset, because we chose to use the same
  typical outburst evolution and jet parameters for both datasets (see
  text).}
\end{center}
\end{table}

\subsection{Jet Properties}

The jet parameters derived for Cyg~X-3 are given in Table 1.  We find
a value of $k$ of 2.54, which is almost exactly what one would expect
for an adiabatic jet flow $k_{ad}=2(s+2)/3\approx 2.51$. The derived
value for $b$ suggests that the magnetic field is rather randomly
oriented than purely perpendicular ($b=1.0$) or parallel ($b=2.0$) to
the jet.  The fact that the value of $r$ is found greater than 1
suggests that the jet opening angle increases with distance from the
core. This was previously found for the other microquasar GRS~1915+105
analysed with the same method and is opposite to what we found for the
quasars 3C~273 and 3C~279.  If the microquasar jets are indeed
trumpet-like (this was suggested already by Marti et al. 1992), it
could be the result from the pressure gradient of material surrounding
the jet or from diverging external magnetic field lines channelling
the jet plasma. A high value of $r$ could also be the signature of a
decelerating jet flow, but as for GRS~1915+105 we do not find a better
fit by relaxing the value of $d$ to prove this hypothesis. As the
effect of the high $r$ value is only to fasten the decrease of the
spectral turnover frequency with time, T\"urler \& Lindfors (2007)
proposed an alternative origin for this behaviour.  They discuss the
possibility that the observed spectral turnover might not be the
self-absorption turnover, as assumed here, but be the result of a
low-energy cut-off in the electron spectrum. If the characteristic
synchrotron frequency associated to the electrons with minimal energy
would be higher than the synchrotron self-absorption frequency during
the early stages of the outburst's evolution, we would indeed have a
faster evolution of the spectral turnover with time. However, further
investigation of this alternative to the trumpet-like jet explanation
is beyond the scope of this paper.

We find that in order to fit the data well, the electron index $s$
needs to be below 2.0, our best fit model having $s=1.77$. Such low
$s$ values have been suggested for several AGNs by Valtaoja et
al. (1988), Hughes et al. (1989, 1991) and Stevens et al. (1995, 1996)
although traditionally such low $s$ values are disfavoured by theory as
for $s<2$ the strong radiative losses of the dominant high-energy
electrons would lead to a substantial pressure decrease along the jet
and prevent the shock to propagate far (Marscher \& Gear 1985).

However, recent theoretical work has shown that an index of $s<2.0$
can be explained with e.g. a high minimum energy of the electron
energy spectrum (Katarzynski et al. 2006), stochastic acceleration
(Virtanen \& Vainio 2005) or a converter mechanism in the radiation
dominated phase (Stern 2003). The hard spectral index can also be
obtained by amplification of particle-accelerating turbulence by the
shock itself. Depending on the magnetic field strength, the
transmission of turbulence through the shock can lead to increased
acceleration efficiency and spectral indices significantly harder than
$s=2$ (see Vainio et al. 2003, 2005, Tammi \& Vainio 2006). Using this
approach, one can calculate that in order to get an index of $s=1.77$
from a simple parallel shock travelling with a speed of $0.63\,c$
(MJ04), a field of $B_{\rm H}\approx0.2\,{\rm G}$ for a hydrogen
plasma and of $B_{\rm pair}\approx6.8\,{\rm mG}$ for a pair plasma
would be enough to allow sufficient amplification. The value for a
pair plasma is well below the upper limit \textbf{$B<0.15\,{\rm G}$}
deduced by MJ04, but the derived magnetic field strength for a
hydrogen plasma is above the upperlimit (see Tammi 2007).

\begin{table*}
\begin{center}
\begin{tabular}{@{}l l l l c c c c c c c c c@{}}\hline \hline
dataset& ref.& $\beta$& $\theta$& $D$& $\Delta l$& $\theta_{\mathrm{src}}$& $B$& $K$& $u_{\mathrm{e}}$& $u_B/u_{\mathrm{e}}$& $E_{\mathrm{tot}}$& $E_{\mathrm{rad}}$\\
    &     &     &[$^\circ$]       &    & [AU]& [mas]& [G]& [erg$^{s-1}$\ cm$^{-3}$]& [erg\ cm$^{-3}$]&             & [erg]           & [erg] \\ \hline
1994& M01 & 0.81& $14$  & 2.7&733 & 6.1& 3.23  & 3.7$\cdot10^{-6}$& 3.2$\cdot10^{-6}$& 1.3$\cdot10^{5}$& 1.7$\cdot10^{44}$& 5.7$\cdot10^{39}$\\
    & MJ04& 0.63& $10.5$& 2.0&321 & 2.7& 0.088 & 1.7$\cdot10^{-2}$& 2.4$\cdot10^{-2}$& 1.3$\cdot10^{-2}$& 8.2$\cdot10^{41}$& 7.6$\cdot10^{39}$\\ \hline
2001& M01 & 0.81& $14$  & 2.7&1162& 9.7& 0.66  & 4.6$\cdot10^{-5}$& 4.8$\cdot10^{-5}$& 3.6$\cdot10^{2}$& 2.8$\cdot10^{43}$& 5.1$\cdot10^{40}$\\
    & MJ04& 0.63& $10.5$& 2.0&508 & 4.3& 0.018 & 2.2$\cdot10^{-1}$& 3.6$\cdot10^{-1}$& 3.6$\cdot10^{-5}$& 4.9$\cdot10^{43}$& 6.8$\cdot10^{40}$\\ \hline
\end{tabular}
\caption{Physical jet properties derived for the average outburst at the point where it reaches its maximum flux density (see text).}
\end{center}
\end{table*}

As already attempted in T\"urler \& Lindfors (2007), it is in
principle possible to derive the physical properties of the jet at the
time when the synchrotron spectrum of the average outburst reaches its
maximum. In addition to the peaking flux density and peaking
frequency, the distance of the source, the jet speed ($\beta$) and the
jet angle to the line of sight ($\theta$) are needed to calculate the
physical properties of the jet.  Unfortunately for Cyg~X-3 these three
quantities have large uncertainties, but in Table~3 the derived values
are given assuming a distance of 10~kpc and making two different
assumptions for $\beta$ and $\theta$ following M01 and MJ04. The
Doppler factor $D$ of the flow is calculated from $\beta$ and
$\theta$. The angular size $\theta_{\mathrm{src}}$ is assumed to be
equal to the width of the jet and is calculated from the distance $\Delta l$
travelled in $\Delta t_{\mathrm{obs}}$ which is the time interval
needed for the outburst to reach its maximum flux density. We assume
here a jet opening half-angle of 2.4$^{\circ}$ (MJ04). The magnetic
field strength $B$, the normalization of the electron energy spectrum
$K$ and the energy density of the relativistic electrons
$u_{\mathrm{e}}$, are then calculated using Eqs. (3) to (5) of
Marscher (1987).

The physical parameters are different for the two different epochs as
the average outburst's peak flux, peak frequency and peak time are
different (see Table~1). The differences between the epochs are
however much smaller than the differences introduced by assuming
different jet flow speeds and viewing angles, which affect the source
angular size $\theta_{\mathrm{src}}$. The physical quantities are
extremely dependent on $\theta_{\mathrm{src}}$:
$B\propto\theta_{\mathrm{src}}^4$,
$K\propto\theta_{\mathrm{src}}^{-2s+5}$,
$u_{\mathrm{e}}\propto\theta_{\mathrm{src}}^{-9}$ and
$u_B/u_{\mathrm{e}}\propto\theta_{\mathrm{src}}^{17}$.  
The parameter values derived using the MJ04 inputs are probably more
trustworthy for the 2001 dataset, as they refer to the same flaring
period.  They can be compared to the values derived in
MJ04. $\theta_{\mathrm{src}}$ derived here is smaller than the sizes
of the 22~GHz VLBA knots which results in a lower estimate of the
magnetic field. We note however that this difference just reflects the
fact that we are not evaluating the source properties at the same time
after the outburst.  We also note that the value derived for the 2001
dataset with the parameters of MJ04 is in good agreement with our
estimation of the magnetic field derived from the index of the
electron energy distribution $s$ in the assumption of a pair
plasma. Another value we can compare to values derived in MJ04 is the
energy density of the relativistic electrons $u_{\mathrm{e}}$, and our
value is above the lower limit of 0.1 erg\ cm$^{-3}$ of MJ04.

To test whether our results are realistic we also derive the
energetics of the average outburst. We calculate the total energy
$E_{\mathrm{tot}}$ in the form of both magnetic field $u_B$ and
relativistic electrons $u_{\mathrm{e}}$ contained in a spherical
source of angular size $\theta_{\mathrm{src}}$.  We find values in the
$\sim$\,10$^{42}$--10$^{44}$\,erg range, which would require a
reasonable energy injection time from about 1 hour to 3 days at a rate
corresponding to the Eddington luminosity of an object of 3 solar
masses. This calculation implicitly assumes that protons do not
contribute significantly to the energetics. In the hypothesis of a jet
plasma made of hydrogen rather than electron-positron pairs, the
protons could carry 2--3 orders of magnitude more energy than the
electrons. This will not significantly affect the total energy
requirement when it is dominated by the magnetic field ($u_{B}\gg
u_{\mathrm{e}}$), but can strongly increase the quoted values when the
particle energy dominates ($u_{B} \ll u_{\mathrm{e}}$). However,
because the balance between $u_{\mathrm{B}}$ and $u_{\mathrm{e}}$ is
extremely sensitive to the source size, it is very difficult to
disfavour a hadronic jet based on such energetic arguments.

Another test is to compare the total energy in the source $E_{\mathrm{tot}}$
with the total radiated energy $E_{\mathrm{rad}}$ of the average outburst. This
quantity is obtained by integrating the outburst flux over time and frequency in
the range shown in Fig.  3b. As expected, we find that the radiated energy
through synchrotron emission is well below the total energy content of the
source. For the 2001 dataset the radiated energy $E_{\mathrm{rad}}$ in this
frequency range is about 1000 times lower than the total energy
$E_{\mathrm{tot}}$ suggesting that the main part of the injected energy will not
be radiated, but will feed the adiabatic expansion of the source.

Because many of the quantities in Table~3 very strongly depend on
the source size, their estimated values differ by several magnitudes.
Nevertheless, we demonstrate here that if the distance, flow speed, viewing
angle and jet opening radius were known with better accuracy, our fitting results
could be used for deriving the other physical parameters of the
jet.

\section{Summary and Conclusion}
We have studied the synchrotron flaring behaviour of Cyg~X-3 during an
extended flaring period in February--March 1994 and a major
outburst in September 2001. By decomposing the multiwavelength
lightcurves into a series of outbursts, we find that during both
epochs the flaring behaviour can be described by a modified
shock-in-jet model.


We derived the average evolution of the outburst during the two  flaring
epochs as well as the characteristics of individual outbursts.  We find that the
outbursts peaking at higher frequency are shorter than the  outbursts peaking at
lower frequencies. We also find that the 2001 outbursts  differ mainly in
peaking flux from the average outburst in the 1994 period. We also derive
parameters describing the jet properties. We find that the index of the electron
energy distribution is hard with $s<2$. 
The best fit suggests the jet geometry to be
trumpet-like, rather than conical, but an alternative explanation to the
observed behaviour is also discussed. Finally, we show that the method can be 
used for deriving the physical properties in the jet of Cyg~X-3 and
discuss the energetics of the average outburst. 

The method originally developed for describing the outbursting behaviour
of quasars has proven to be useful also for microquasar studies. The method is
complementary to radio interferometric jet imaging for deriving the
parameters of relativistic jets. To take full advantage of the capabilities of
this method, densely sampled lightcurves at radio and millimetre, as well as
submillimetre and infrared frequencies are required.

\begin{acknowledgements}
This research has been supported by the Academy of Finland grants
74886 and 80450, Jenny and Antti Wihuri Foundation and V\"ais\"al\"a
Foundation. EJL wishes to thank Linnea Hjalmarsdotter for useful
discussions. DCH gratefully acknowledges support from the Academy of
Finland.  SAT is grateful for support of the Russian Foundation  Base
Research (grant N 05-02-17556). We also thank the anonymous referee
for constructive criticism of the earlier version of this paper.
\end{acknowledgements}

\end{document}